\documentclass[a4paper,12pt]{article}
\def\int {\intop \limits}

\def\fnote#1{\footnote}
\usepackage{graphics}

\begin{document}

\title{Coherent and incoherent radiation from high-energy electron
and the LPM effect in oriented single crystal}
\author{V. N. Baier
and V. M. Katkov\\
Budker Institute of Nuclear Physics,\\ Novosibirsk, 630090, Russia}

\maketitle

\begin{abstract}

The process of radiation from high-energy electron in oriented
single crystal is considered using the method which permits
inseparable consideration of both coherent and incoherent mechanisms
of photon emission. The total intensity of radiation is calculated.
The theory, where the energy loss of projectile has to be taken into
account, agrees quite satisfactory  with available CERN data. It is
shown that the influence of multiple scattering on radiation process
is suppressed due to action of crystal field.

\end{abstract}

\newpage

Recently authors developed a new approach to analysis of pair
creation by a photon in oriented crystals \cite{BK0}. This approach
not only permits to consider simultaneously both the coherent and
incoherent mechanisms of pair creation by a photon but also gives
insight on the Landau-Pomeranchuk-Migdal (LPM) effect (influence of
multiple scattering) on the considered mechanism of pair creation.
In the approach the polarization tensor of photon was used which
includes influence of both external field and multiple scattering of
electrons and positrons in a medium \cite{BK}. In the present paper
the analysis of process of radiation from a high-energy electron in
oriented crystal includes influence of both an external field and
the multiple scattering of electron. This makes possible indivisible
consideration of both coherent and incoherent mechanisms of photon
emission as well as analysis of influence of the LPM effect on
radiation process.

The properties of radiation are connected directly with details of
motion of emitting particle. The momentum transfer from a particle
to a crystal we present in a form ${\bf q}=<{\bf q}>+{\bf q}_s$,
where $<{\bf q}>$ is the mean value of momentum transfer calculated
with averaging over thermal(zero) vibrations of atoms in a crystal.
The motion of particle in an averaged potential of crystal, which
corresponds to the momentum transfer $<{\bf q}>$, determines the
coherent mechanism of radiation. The term ${\bf q}_s$ is attributed
to the random collisions of particle which define the incoherent
radiation. Such random collisions we will call "scattering" since
$<{\bf q}_s>=0$. If the radiation formation length is large with
respect to distances between atoms forming the axis, the additional
averaging over the atom position should be performed.

Under some generic assumptions the general theory of the coherent
radiation mechanism was developed in \cite{BKS1}. If the electron
angle of incidence $\vartheta_0$ (the angle between electron
momentum {\bf p} and the axis (or plane)) is small $\vartheta_0 \ll
V_0/m$, where $V_0$ is the characteristic scale of the potential,
the field $E$ of the axis (or plane) can be considered constant over
the pair formation length and the constant-field approximation
(magnetic bremsstrahlung limit) is valid. In this case the behavior
of radiation probability is determined by the parameter
\begin{equation}
\chi=\frac{\varepsilon}{m}\frac{E}{E_0}, \label{1}
\end{equation}
where $\varepsilon$ is the electron energy, $m$ is the electron
mass, $E_0=m^2/e=1.32 \cdot 10^{16}$ V/cm is the critical field, the
system $\hbar=c=1$ is used. The very important feature of coherent
radiation mechanism is the strong enhancement of its probability at
high energies (from factor $\sim 10$ for main axes in crystals of
heavy elements like tungsten to factor $\sim 170$ for diamond)
comparing with the Bethe-Heitler mechanism which takes place in an
amorphous medium. If $\vartheta_0 \gg V_0/m$ the theory passes over
to the coherent bremsstrahlung theory (see \cite{D},\cite{T}
\cite{BKS}). Side by side with coherent mechanism the incoherent
mechanism of radiation is acting. In oriented crystal this mechanism
changes also with respect to an amorphous medium \cite{BKS2}. The
details of theory and description of experimental study of radiation
which confirms the mentioned enhancement can be found in \cite{BKS}.
The study of radiation in oriented crystals is continuing and new
experiments are performed recently \cite{KMU}, \cite{BKi}.

At high energies the multiple scattering of radiating electron (the
LPM effect) suppresses radiation probability when $\varepsilon \geq
\varepsilon_e$. In an amorphous medium (or in crystal in the case of
random orientation) the characteristic electron energy starting from
which the LPM effect becomes essential is $\varepsilon_e \sim
2.5$~TeV for heavy elements \cite{BK1} and this value is inversely
proportional to the density. In the vicinity of crystalline axis
(just this region gives the crucial contribution to the
Bethe-Heitler mechanism) the local density of atoms is much higher
than average one and for heavy elements and at low temperature the
gain could attain factor $\sim 10^3$. So in this situation the
characteristic electron energy can be $\varepsilon_0 \sim 2.5$~GeV
and this energy is significantly larger than "threshold" energy
$\varepsilon_t$ starting from which the probability of coherent
radiation exceeds the incoherent one. It should be noted that the
main contribution into the multiple scattering gives the small
distance from axis where the field of crystalline axis attains the
maximal value. For the same reason the LPM effect in oriented
crystals originates in the presence of crystal field and
nonseparable from it. This means that in problem under consideration
we have both the dense matter with strong multiple scattering and
high field of crystalline axis.

Below we consider case $\vartheta_0 \ll V_0/m$. Than the distance of
an electron from axis $\mbox{\boldmath$\varrho$}$ as well as the
transverse field of the axis can be considered as constant over the
formation length. For an axial orientation of crystal the ratio of
the atom density $n(\varrho)$ in the vicinity of an axis to the mean
atom density $n_a$ is
\begin{equation}
\frac{n(x)}{n_a}=\xi(x)=\frac{x_0}{\eta_1}e^{-x/\eta_1},\quad
\varepsilon_0=\frac{\varepsilon_e}{\xi(0)},
\label{2}
\end{equation}
where
\begin{equation}
x_0=\frac{1}{\pi d n_a a_s^2}, \quad  \eta_1=\frac{2
u_1^2}{a_s^2},\quad x=\frac{\varrho^2}{a_s^2}, \label{3}
\end{equation}
Here $\varrho$ is the distance from axis, $u_1$ is the amplitude of
thermal vibration, $d$ is the mean distance between atoms forming
the axis, $a_s$ is the effective screening radius of the axis
potential (see Eq.(9.13) in \cite{BKS})
\begin{equation}
U(x)=V_0\left[\ln\left(1+\frac{1}{x+\eta} \right)-
\ln\left(1+\frac{1}{x_0+\eta} \right) \right]. \label{4}
\end{equation}
The local value of parameter $\chi(x)$ (see Eq.(\ref{1})) which
determines the radiation probability in the field Eq.(\ref{4}) is
\begin{equation}
\chi(x)=-\frac{dU(\varrho)}{d\varrho}\frac{\varepsilon}{m^3}=\chi_s
\frac{2\sqrt{x}}{(x+\eta)(x+\eta+1)},\quad \chi_s=\frac{V_0
\varepsilon}{m^3a_s}\equiv \frac{\varepsilon}{\varepsilon_s}.
 \label{5}
\end{equation}
The parameters of the axial potential for the ordinarily used
crystals are given in Table 9.1 in \cite{BKS}. The particular
calculation below will be done for tungsten crystals studied in
\cite{KMU}. The relevant parameters are given in Table 1. It is
useful to compare the characteristic energy $\varepsilon_0$ with
"threshold" energy $\varepsilon_t$ for which the radiation intensity
in the axis field becomes equal to the Bethe-Maximon one. Since the
maximal value of parameter $\chi(x)$:
\begin{equation}
\chi_m=\chi(x_m),\quad
x_m=\frac{1}{6}(\sqrt{1+16\eta(1+\eta)}-1-2\eta),\quad
\chi_m=\frac{\varepsilon}{\varepsilon_m}
\label{6}
\end{equation}
is small for such electron energy $(\varepsilon_t \ll
\varepsilon_m)$, one can use the decomposition of radiation
intensity over powers of $\chi$ (see Eq.(4.52) in \cite{BKS}) and
carry out averaging over $x$. Retaining three terms of decomposition
we get
\begin{eqnarray}
&&I^F = \frac{8\alpha m^2
\chi_s^2}{3x_0}\left(a_0(\eta)-a_1(\eta)\chi_s +
a_2(\eta)\chi_s^2+\ldots \right),
\nonumber \\
&& a_0(\eta)=(1+2\eta)\ln \frac{1+\eta}{\eta}-2,
\nonumber \\
&& a_1(\eta)=\frac{165 \sqrt{3}
\pi}{64}\left[\frac{1}{\sqrt{\eta}}-\frac{1}{\sqrt{1+\eta}}-
4\left(\sqrt{1+\eta} - \sqrt{\eta} \right)^3\right],
\nonumber \\
&& a_2(\eta)=64\left[(1+2\eta)\left(\frac{1}{\eta(1+
\eta)}+30\right)- 12(1+5\eta(1+\eta))\ln \frac{1+\eta}{\eta}\right].
\label{7}
\end{eqnarray}

The intensity of incoherent radiation in low energy region
$\varepsilon \leq \varepsilon_t \ll \varepsilon_m$ is (see
Eq.(21.16) in \cite{BKS} and Eq.(\ref{a.16}) in Appendix A)
\begin{eqnarray}
&&I^{inc}=\frac{\alpha
m^2}{4\pi}\frac{\varepsilon}{\varepsilon_e}g_0\left[1+34.4
\left(\overline{\chi^2\ln \chi}+ 2.54\overline{\chi^2}\right)\right]
\nonumber \\
&&
g_0=1+\frac{1}{L_0}\left[\frac{1}{18}-h\left(\frac{u_1^2}{a^2}\right)\right],
\quad \overline{f} = \int_0^{\infty} f(x)
e^{-\frac{x}{\eta_1}}\frac{dx}{\eta_1},
\label{8}
\end{eqnarray}
where
\begin{eqnarray}
&& \varepsilon_e=\frac{m}{16\pi Z^2\alpha^2\lambda_c^3n_aL_0},\quad
L_0=\ln(ma)+ \frac{1}{2}-f(Z\alpha),
\nonumber \\
&&h(z)=-\frac{1}{2}\left[1+(1+z)e^{z}{\rm Ei}(-z) \right],\quad
a=\frac{111Z^{-1/3}}{m},
\nonumber \\
&& f(\xi)={\rm Re} \left[\psi(1+i\xi)-\psi(1)
\right]=\sum_{n=1}^{\infty} \frac{\xi^2}{n(n^2+\xi^2)}, \label{9}
\end{eqnarray}
here $\psi(z)$ is the logarithmic derivative of the gamma function,
Ei($z$) is the integral exponential function, $f(\xi)$ is the
Coulomb correction. For $\chi=0$ this intensity differs from the
Bethe-Maximon intensity only by the term $h(u_1^2/a^2)$ which
reflects the nongomogeneity of atom distribution in crystal. For
$u_1\ll a$ one has $h(u_1^2/a^2) \simeq
-(1+C)/2+\ln(a/u_1),~C=0.577..$ and so this term characterizes the
new value of upper boundary of impact parameters $u_1$ contributing
to the value $<{\bf q}_s^2>$ instead of screening radius $a$ in an
amorphous medium.

Conserving in Eq.(\ref{7}) only the main (the first) term of
decomposition, which corresponds to the classical radiation
intensity, neglecting the corrections in Eq.(\ref{8})
($g_0=1,~\chi=0$), using the estimate $V_0 \simeq Z\alpha/d$ and
Eqs.(\ref{3}), (\ref{5}), we get
\begin{equation}
\varepsilon_t \simeq \frac{3 L_0 d m^2}{2\pi a_0(\eta)} =
63\frac{L_0 d}{a_0(\eta)}{\rm MeV},
\label{10}
\end{equation}
where the distance $d$ is taken in units $10^{-8}$ cm. Values of
$\varepsilon_t$ found using this estimate for tungsten, axis
$<111>$, $d$=2.74 $\cdot 10^{-8}$ cm are consistent with points of
intersection of coherent and incoherent intensities in Fig.1 (see
Table 1). For some usable crystals (axis $<111>$, room temperature)
one has from Eq.(\ref{10})
\begin{equation}
\varepsilon_t({\rm C_{(d)}}) \simeq 0.47~ {\rm GeV},\quad
\varepsilon_t({\rm Si}) \simeq 2.0~ {\rm GeV},\quad
\varepsilon_t({\rm Ge}) \simeq 1.7~ {\rm GeV},
\label{11}
\end{equation}
so this values of $\varepsilon_t$ are somewhat larger than in
tungsten except the diamond very specific crystal where value of
$\varepsilon_t$ is close to tungsten one.

For large values of the parameter $\chi_m~(\varepsilon \gg
\varepsilon_m)$ the incoherent radiation intensity is suppressed due
to the action of the axis field. In this case the local intensity of
radiation can by written as (see Eq.(7.129) in \cite{BKS})
\begin{equation}
I^{inc}=\frac{29
\Gamma(1/3)}{3^{1/6}2430}\frac{\varepsilon}{\varepsilon_e}\frac{\alpha
m^2}{\chi^{2/3}(x)}\left[g_0+\frac{1}{L_0}\left( 0.727+\frac{\ln
\chi(x)}{3} \right)\right]. \label{12}
\end{equation}
Here we take into account that
\begin{equation}
\ln \frac{1}{\gamma \vartheta_1}=\ln(ma) \rightarrow
\ln(ma)-h\left(\frac{u_1^2}{a^2}\right)-f(Z\alpha)=L_0-\frac{1}{2}.
\label{12a}
\end{equation}
 Averaging the function $(\chi(x))^{-2/3}$ and
$\ln\chi(x)(\chi(x))^{-2/3}$ over $x$ according with Eq.(\ref{8})
one can find the effective value of upper boundary of the transverse
momentum transfer ($\propto m\chi^{1/3}_m$ instead of $m$) which
contributes to the value $<{\bf q}_s^2>$. Using the obtained results
we determine the effective logarithm $L$ by means of interpolation
procedure
\begin{equation}
L=L_0g,\quad g=g_0+\frac{1}{6 L_0}\ln \left(1+70\chi_m^2\right).
\label{13}
\end{equation}
Let us introduce the local characteristic energy (see Eq.(\ref{2}))
\begin{equation}
\varepsilon_c(x)=
\frac{\varepsilon_e(n_a)}{\xi(x)g}=\frac{\varepsilon_0}{g}e^{x/\eta_1},
\label{14}
\end{equation}
In this notations the contribution of multiple scattering into the
local intensity for small values of $\chi_m$ and
$\varepsilon/\varepsilon_0$ has a form (see Eq.(15) in \cite{BK3})
\begin{equation}
I^{LPM}(x)=-\frac{\alpha m^2}{4\pi}
\frac{\varepsilon}{\varepsilon_c(x)}\left[\frac{4\pi\varepsilon}{15
\varepsilon_c(x)}+\frac{64 \varepsilon^2}{21
\varepsilon_c^2(x)}\left(\ln
\frac{\varepsilon}{\varepsilon_c(x)}+2.04\right)\right]. \label{15}
\end{equation}
Integrating this expression over $x$ with the weight $1/x_0$ we get
\begin{equation}
I^{LPM}=\frac{\alpha m^2}{4\pi}
\frac{\varepsilon}{\varepsilon_e}g\left[-\frac{2\pi\varepsilon g}{15
\varepsilon_0}+\frac{64}{63}\frac{\varepsilon^2 g^2}{
\varepsilon_0^2}\left(\ln \frac{\varepsilon_0}{\varepsilon
g}-1.71\right)\right]. \label{16}
\end{equation}
It should be noted that found Eq.(\ref{16}) has a good accuracy only
for energy much smaller (at least on one order of magnitude) than
$\varepsilon_0$ (see discussion after Eq.(15) in \cite{BK3}).

The spectral probability of radiation under the simultaneous action
of multiple scattering and an external constant field was derived in
\cite{BKS} (see Eqs.(7.89) and (7.90)). Multiplying the expression
by $\omega$ and integrating over $\omega$ one obtains the total
intensity of radiation $I$. For further analysis and numerical
calculation it is convenient to carry out some transformations
\begin{enumerate}
\item Changing of variables:~$\nu \rightarrow a\nu/2,~ \tau \rightarrow
2t/a,~(\nu\tau \rightarrow \nu t)$.
\item
Turn the contour of integration over $t$ at the angle $-\pi/4$.
\end{enumerate}
One finds after substitution $t \rightarrow \sqrt{2}t$
\begin{eqnarray}
&& I(\varepsilon)=\frac{\alpha m^2}{2\pi}\int_0^1 \frac{y dy}{1-y}
\int_0^{x_0}\frac{dx}{x_0}G_r(x, y),\quad G_r(x, y)=\int_0^{\infty}
F_r(x, y, t)dt -r_3\frac{\pi}{4},
\nonumber \\
&& F_r(x, y, t)={\rm Im}\left\lbrace
e^{\varphi_1(t)}\left[r_2\nu_0^2
(1+ib_r)\varphi_2(t)+r_3\varphi_3(t) \right] \right\rbrace,\quad
b_r=\frac{4\chi^2(x)}{u^2\nu_0^2},
\nonumber \\
&& y=\frac{\omega}{\varepsilon}, \quad u=\frac{y}{1-y},\quad
\varphi_1(t)=(i-1)t+b_r(1+i)(\varphi_2(t)-t),
\nonumber \\
&&
\varphi_2(t)=\frac{\sqrt{2}}{\nu_0}\tanh\frac{\nu_0t}{\sqrt{2}},\quad
\varphi_3(t)=\frac{\sqrt{2}\nu_0}{\sinh(\sqrt{2}\nu_0t)},
\label{17}
\end{eqnarray}
where
\begin{equation}
r_2=1+(1-y)^2,\quad r_3=2(1-y),\quad \nu_0^2=\frac{1-y}{y}
\frac{\varepsilon}{\varepsilon_c(x)},
\label{18}
\end{equation}
$\omega$ is the photon energy, the function $\varepsilon_c(x)$ is
defined in Eq.(\ref{14}) and $\chi(x)$ is defined in Eq.(\ref{5}).
The expression for the spectral probability of radiation used in the
above derivation can be found from the spectral form of Eq.(16) in
\cite{BK0} ($dW/dy=\omega dW/d\varepsilon $) using the standard QED
substitution rules: $\varepsilon \rightarrow -\varepsilon,~\omega
\rightarrow -\omega,~\varepsilon^2d\varepsilon \rightarrow
\omega^2d\omega$ and exchange $\omega_c(x) \rightarrow
4\varepsilon_c(x)$.

The inverse radiation length in tungsten crystal (axis $<111>$)
$1/L^{cr}(\varepsilon)=I(\epsilon)/\varepsilon$ Eq.(\ref{17}), well
as coherent contribution
$1/L^F(\varepsilon)=I^F(\varepsilon)/\varepsilon$ Eq.(\ref{19}) and
incoherent contribution
$1/L^{inc}(\varepsilon)=I^{inc}(\varepsilon)/\varepsilon$
Eq.(\ref{21})  are shown in Fig.1 for two temperatures T=100 K and
T=293 K as a function of incident electron energy $\varepsilon$. In
low energy region ($\varepsilon \leq 0.3$~GeV) the asymptotic
expressions Eqs.(\ref{7}) and (\ref{8}) are valid. One can see that
at temperature T=293 K the intensity $I^F(\varepsilon)$ is equal to
$I^{inc}(\varepsilon)$ at $\varepsilon \simeq 0.4$~GeV and
temperature T=100 K the intensity $I^F(\varepsilon)$ is equal to
$I^{inc}(\varepsilon)$ at $\varepsilon \simeq 0.7$~GeV. The same
estimates follow from comparison of Eqs.(\ref{7}) and (\ref{8}), see
also Eq.(\ref{10}). At higher energies the intensity
$I^F(\varepsilon)$ dominates while the intensity
$I^{inc}(\varepsilon)$ decreases monotonically.

The inverse radiation length given in Fig.1 can be compared with
data directly only if the crystal thickness $l \ll
L^{cr}(\varepsilon)$ (thin target). Otherwise one has to take into
account the energy loss. The corresponding analysis is simplified
essentially if $l \leq L^{min}=({\rm max}
(I(\varepsilon)/\varepsilon))^{-1}$. The radiation length
$L^{cr}(\varepsilon)$ varies slowly on the electron trajectory for
such thicknesses. This is because of weak dependence of
$L^{cr}(\varepsilon)$ on energy in the region $L^{cr}(\varepsilon)
\simeq L^{min}$ and the relatively large value of
$L^{cr}(\varepsilon) \gg L^{min}$ in the region where this
dependence is essential but variation of energy on the thickness $l$
is small. For W, axis $<111>$, T=293 K one has $L^{min}=320~\mu m$
at energy $\varepsilon=300$~GeV, see Fig.1. For this situation
dispersion can be neglected (see discussion in Sec.17.5 of
\cite{BKS}) and energy loss equation acquires the form
\begin{equation}
\frac{1}{\varepsilon}\frac{d\varepsilon}{dl}=
-L^{cr}(\varepsilon)^{-1}\equiv -\frac{I(\varepsilon)}{\varepsilon}.
\label{18a}
\end{equation}
In the first approximation the final energy of electron is
\begin{equation}
\varepsilon_1=\varepsilon_0
\exp\left(-l/L^{cr}(\varepsilon_0)\right), \label{18b}
\end{equation}
where $\varepsilon_0$ is the initial energy. In the next
approximation one has
\begin{equation}
\ln\frac{\varepsilon(l)}{\varepsilon_0}=-L^{cr}(\varepsilon_0)
\int_{\varepsilon_1}^{\varepsilon_0}L^{cr}(\varepsilon)^{-1}
\frac{d\varepsilon}{\varepsilon}. \label{18c}
\end{equation}
If the dependence of $L^{cr}(\varepsilon)^{-1}$ on $\varepsilon$ is
enough smooth it's possible to substitute the function
$L^{cr}(\varepsilon)^{-1}$ by an average value with the weight
$1/\varepsilon$:
\begin{equation}
L^{cr}(\varepsilon)^{-1} \rightarrow
\frac{\varepsilon_0L^{cr}(\varepsilon_1)^{-1}+\varepsilon_1L^{cr}(\varepsilon_0)^{-1}}
{\varepsilon_0+\varepsilon_1}\equiv \frac{1}{\overline{L}}.
\label{18d}
\end{equation}
Numerical test confirms this simplified procedure. Using it we find
\begin{equation}
\ln\frac{\varepsilon(l)}{\varepsilon_0}=
-\frac{L^{cr}(\varepsilon_0)}{\overline{L}}
\ln\frac{\varepsilon_0}{\varepsilon_1}=-\frac{l}{\overline{L}},\quad
\frac{\Delta
\varepsilon}{\varepsilon_0}=1-\exp\left(-\frac{l}{\overline{L}}\right)
\equiv \frac{l}{L^{ef}}. \label{18e}
\end{equation}

Enhancement of  radiation length (the ratio of Bethe-Maximon
radiation length $L^{BM}$ and $L^{ef}$ ) in tungsten, axis $<111>$,
T=293 K is shown in Fig.2. The curve 1 is for the target with
thickness $l=200~\mu m$, where the energy loss was taken into
account according using the simplified procedure Eq.(\ref{18e}). The
curve 2 is for a considerably more thinner target, where one can
neglect the energy loss. The only available data are from
\cite{KMU}. The measurement of radiation from more thin targets is
of evident interest.

In order to single out the influence of the multiple scattering (the
LPM effect) on the process under consideration, we should consider
both the coherent and incoherent contributions. The probability of
coherent radiation is the first term ($\nu_0^2=0$) of the
decomposition of Eq.(\ref{17}) over $\nu_0^2$.  The coherent
intensity of radiation is (compare with Eq.(17.7) in \cite{BKS})
\begin{equation}
I^{F}(\varepsilon)=\int_0^{x_0}I(\chi)\frac{dx}{x_0}. \label{19}
\end{equation}
Here $I(\chi)$ is the radiation intensity in constant field
(magnetic bremsstrahlung limit, see Eqs. (4.50), (4.51) in
\cite{BKS}). It is convenient to use the following representation
for $I(\chi)$
\begin{eqnarray}
&&I(\chi)=i\frac{\alpha m^2}{2\pi}
\int_{\lambda-i\infty}^{\lambda+i\infty}
\left(\frac{\chi^2}{3}\right)^s \Gamma\left(1-s\right)
\Gamma\left(3s-1\right)(2s-1)(s^2-s+2)\frac{ds}{\cos\pi s},
\nonumber \\
&&\frac{1}{3}<\lambda<1.
\label{20}
\end{eqnarray}

The intensity of incoherent radiation is the second term ($\propto
\nu_0^2$) of the mentioned decomposition. In Appendix A the new
representation of this intensity is derived, which is suitable for
both analytical and numerical calculation:
\begin{equation}
I^{inc}(\varepsilon)=\frac{\alpha m^2}{60\pi}
\frac{\varepsilon}{\varepsilon_0}g\int_0^{x_0}e^{-x/\eta_1}J(\chi)\frac{dx}{x_0},
\label{21}
\end{equation}
where $J(\chi)$ is defined in Eq.(\ref{a.14}).

The contribution of the LPM effect in the total intensity of
radiation $I$ Eq.(\ref{17}) is defined as
\begin{equation}
I^{LPM}=I - I^F -I^{inc} \label{23}
\end{equation}
The relative contribution (negative since the LPM effect suppresses
the radiation process) $\Delta=-I^{LPM}/I$ is shown in Fig.3. This
contribution has the maximum $\Delta \simeq 0.8$\% at $\varepsilon
\simeq 0.7$~GeV for T=293 K and $\Delta \simeq 0.9$\% at
$\varepsilon \simeq 0.3$~GeV for T=100 K or, in general, at
$\varepsilon \sim \varepsilon_t$. The left part of the curves is
described quite satisfactory by  Eq.(\ref{16}). For explanation of
right part of the curves let us remind that at $\varepsilon \gg
\varepsilon_m$ the behavior of the radiation intensity at $x \sim
\eta_1$ is defined by the ratio of the contributions to the momentum
transfer of multiple scattering and that of the external field on
the formation length $l_f$ (see Eq.(21.3) in \cite{BKS})
\begin{eqnarray}
&&k=\frac{<{\bf q}_s^2>}{<{\bf q}>^2}
=\frac{\dot{\vartheta}_s^2l_f}{(wl_f)^2}\sim
\frac{\varepsilon}{\varepsilon_0}
\chi_m^{-4/3}=\frac{\varepsilon}{\varepsilon_0}
\left(\frac{\varepsilon_m}{\varepsilon}\right)^{4/3},
\nonumber \\
&&\frac{1}{L^F} \sim \frac{\alpha}{l_f} \sim \frac{\alpha
m^2}{\varepsilon}\chi_m^{2/3} = \frac{\alpha
m^2}{\varepsilon_m}\chi_m^{-1/3}, \label{24}
\end{eqnarray}
where $w$ is an acceleration in an external field. The linear over
$k$ term determines the contribution into intensity of incoherent
process: $1/L^{inc}(\varepsilon \gg \varepsilon_m) \sim
k/L^F(\varepsilon) \sim \alpha m^2/(\varepsilon_0 \chi_m^{2/3})$.
The LPM effect is defined by the next term of decomposition over
$k~(\propto k^2)$ and decreases with energy even faster than
$1/L^{inc}(\varepsilon)$. Moreover one has to take into account that
at $\varepsilon \geq \varepsilon_s$ the contribution of relevant
region $x \sim \eta_1$ into the total radiation intensity is small
and $1/L^F(\varepsilon)$ decreases with the energy growth as
$\chi_m^{-1/3}$. For such energies the main contribution gives the
region $x \sim \chi_s^{2/3}=(\varepsilon/\varepsilon_s)^{2/3}$ and
$1/L^{cr}(\varepsilon)$ increases until energy $\varepsilon \sim 10
\varepsilon_s$ (see Fig.1). This results in essential reduction of
relative contribution of the LPM effect $\Delta$.

It's instructive to compare the LPM effect in oriented crystal for
radiation and pair creation processes. The manifestation of the LPM
effect is essentially different because of existence of threshold in
pair creation process. The threshold energy $\omega_m$ is relatively
high (in W, axis $<111>$, $\omega_m \sim 8$~GeV for T=100 K and
$\omega_m \sim 14$~GeV for T=293 K). Below $\omega_m$ influence of
field of axis is weak and the relative contribution of the LPM
effect attains 5.5 \% for T=100 K \cite{BK0}. There is no threshold
in radiation process and $I^F$ becomes larger than  $I^{inc}$ at
much lower energy $\varepsilon_t$ and starting from this energy the
influence of field of axis suppresses strongly the LPM effect. So
the energy interval in which the LPM effect could appear is much
narrower than for pair creation and its relative contribution is
less than 1 \% in W, axis $<111>$. Since value of $\varepsilon_t$
depends weakly on $Z$ (Eq.(\ref{10})), $\varepsilon_m \propto
Z^{-1}$ (Eqs.(\ref{5}), (\ref{6})) and $\varepsilon_0 \propto
Z^{-2}$ (Eq.(\ref{9})) the relative contribution of the LPM effect
$\Delta$ for light elements significantly smaller. Thus, the above
analysis shows that influence of multiple scattering on basic
electromagnetic processes in oriented crystal (radiation and pair
creation) is very limited especially for radiation process.

\vspace{0.5cm}

{\bf Acknowledgments}

We are grateful to U.Uggerhoj for data. The authors are indebted to
the Russian Foundation for Basic Research supported in part this
research by Grant 03-02-16154.

\newpage

\renewcommand \theequation{\thesection.\arabic{equation}}

\setcounter{equation}{0}
\Alph{equation}

\appendix

\section{Appendix}

{\large {\bf New representation of the intensity of the incoherent
radiation in external field, asymptotic expansions}}
\vskip3mm
In
the expression for the intensity of incoherent radiation enters
following integral over photon energy $\omega$~(see Eq.(21.21) in
\cite{BKS}):
\begin{equation}
J(\chi)=\int_0^{1}\left[y^2(f_1(z)+f_2(z))+2(1-y)f_2(z)\right]dy,\quad
z=\left(\frac{y}{\chi(1-y)}\right)^{2/3},
\label{a.1}
\end{equation}
where $y=\omega/\varepsilon$, the functions $f_1(z)$ and $f_2(z)$
are defined in the just mentioned equation in \cite{BKS}:
\begin{eqnarray}
&& f_1(z)=z^4\Upsilon(z)-3z^2\Upsilon'(z)-z^3,
\nonumber \\
&& f_2(z)=(z^4+3z)\Upsilon(z)-5z^2\Upsilon'(z)-z^3,
\label{a.11a}
\end{eqnarray}
here $\Upsilon(z)$ is the Hardy function:
\begin{equation}
\Upsilon(z)=\int_0^{\infty}\sin\left(zt+\frac{t^3}{3}\right)dt
\label{a.12a}
\end{equation}
Introducing the variable $\eta=y/(\chi(1-y))$ we obtain
\begin{eqnarray}
&&
J(\chi)=\int_0^{\infty}\left[\frac{\chi^3\eta^2}{(1+\eta\chi)^2}(f_1+f_2)
+\frac{2\chi}{(1+\eta\chi)}f_2\right]\frac{d\eta}{(1+\eta\chi)^2}
\nonumber \\
&&
=\frac{\chi^3}{6}\frac{d^2}{d\chi^2}(J_1(\chi)+J_2(\chi))+\frac{d}{d\chi}(\chi^2J_2(\chi)),
\label{a.2}
\end{eqnarray}
where
\begin{equation}
J_{1,2}(\chi)=\int_0^{\infty}f_{1,2}(z)\frac{d\eta}{(1+\eta\chi)^2},\quad
z=\eta^{2/3}.
\label{a.3}
\end{equation}
Integrating Eq.(\ref{a.3}) by parts we find
\begin{equation}
J_{1,2}(\chi)=\frac{f_{1,2}(\infty)}{\chi}-\frac{2}{3}\int_0^{\infty}f_{1,2}'(z)
\frac{\eta^{2/3}d\eta}{(1+\eta\chi)}.
\label{a.4}
\end{equation}
Since the integral Eq.(\ref{a.4}) for separate terms of functions
$f_{1.2}'(z)$ diverges, one has to transform it to an another form.
We represent the functions $f_{1,2}'(z)$ in terms of derivative of
the Hardy functions
\begin{equation}
f_{1}'(z)=z^2\Upsilon^{(5)}(z)-3z\Upsilon^{(4)},\quad
f_{2}'(z)=z^2\Upsilon^{(5)}(z)-5z\Upsilon^{(4)}+3\Upsilon^{(3)},
\label{a.5}
\end{equation}
where we used equations
\begin{equation}
z\Upsilon(z)=\Upsilon''(z)+1,\quad
\Upsilon^{(n+3)}(z)=(n+1)\Upsilon^{(n)}+z\Upsilon^{(n+1)}.
\label{a.6}
\end{equation}
Now we will show that
\begin{equation}
\int_0^{\infty}
\eta^{2/3}f_{1,2}'(z)d\eta=\frac{3}{2}\int_0^{\infty}
z^{3/2}f_{1,2}'(z)dz=0. \label{a.7}
\end{equation}
Using Eq.(\ref{a.5}) and integration by parts, one can reduce all
the integrals in Eq.(\ref{a.7}) to the form
\begin{eqnarray}
&&\int_0^{\infty} \Upsilon'(z)\frac{dz}{\sqrt{z}}={\rm
Re}\int_0^{\infty}\frac{dz}{\sqrt{z}}\int_0^{\infty}\tau
\exp\left(iz\tau+\frac{i\tau^3}{3}\right)d\tau
\nonumber \\
&& =\frac{4}{\sqrt{3}}{\rm
Re}\left(\int_0^{\infty}e^{ix^2}dx\right)^2=0.
\label{a.8}
\end{eqnarray}
The last equation permits one to rewrite Eq.(\ref{a.4}) as
\begin{equation}
J_{1,2}(\chi)=\frac{f_{1,2}(\infty)}{\chi}+i_{1,2}(\chi),\quad
i_{1,2}(\chi)=\chi\int_0^{\infty}f_{1,2}'(z)\frac{z^3dz}{1+\chi
z^{3/2}}.
\label{a.9}
\end{equation}

Entering in Eq.(\ref{a.9}) expression $(1+u)^{-1}$ we present as
contour integral
\begin{equation}
\frac{1}{(1+u)}=\frac{i}{2}\int_{\lambda-i\infty}^{\lambda+i\infty}\frac{u^s}{\sin\pi
s}ds,\quad u=\chi z^{3/2},\quad -1<\lambda<0. \label{a.10}
\end{equation}
Using the standard form of the Hardy function one has
\begin{equation}
\Upsilon^{(n)}=\frac{d^n}{dz^n}{\rm Im}\int_0^{\infty}
\exp\left(i\left(z\tau+\frac{\tau^3}{3}\right)\right)d\tau ={\rm
Im}\int_0^{\infty}(i\tau)^n
\exp\left(i\left(z\tau+\frac{\tau^3}{3}\right)\right)d\tau
\label{a.11}
\end{equation}
Substituting in the integral in Eq.(\ref{a.9}) the functions
$f_{1,2}'(z)$ in the form Eqs.(\ref{a.5}), (\ref{a.11}) and
integrating over the variables $z$ and $\tau$ we obtain
\begin{equation}
i_{1,2}(\chi)=\frac{i\pi\chi}{12}\int_{\lambda-i\infty}^{\lambda+i\infty}
\left(\frac{\chi}{\sqrt{3}}\right)^s
\frac{A_{1,2}(s)}{\Gamma\left(1+s/2\right)}\frac{ds}{\sin^2(\pi
s/2)}, \label{a.12}
\end{equation}
where $\Gamma(s)$ is  the gamma function,
\begin{eqnarray}
&&A_{1}(s)=\Gamma\left(\frac{3s}{2}+6\right) +
3\Gamma\left(\frac{3s}{2}+5\right),
\nonumber \\
&&A_{2}(s)=\Gamma\left(\frac{3s}{2}+6\right) +
5\Gamma\left(\frac{3s}{2}+5\right)+3\Gamma\left(\frac{3s}{2}+4\right).
\label{a.13}
\end{eqnarray}
Substituting Eqs.(\ref{a.12}), (\ref{a.13}) into Eq.(\ref{a.9}) and
using Eq.(\ref{a.2}), we get after change of variable $s \rightarrow
2s$, displacement of integration contour and reduction of similar
terms the final expression for $J(\chi)$
\begin{equation}
J(\chi)=\frac{i\pi}{2}\int_{\lambda-i\infty}^{\lambda+i\infty}
\frac{\chi^{2s}}{3^s}
\frac{\Gamma(1+3s)}{\Gamma(s)}R(s)\frac{ds}{\sin^2\pi s},\quad
-\frac{1}{3} < \lambda <0 \label{a.14}
\end{equation}
where
\begin{equation}
R(s)=15+43s+31s^2+28s^3+12s^4.
\label{a.15}
\end{equation}

In the case $\chi \ll 1$, closing the integration contour on the
right, one can calculate the asymptotic series in powers of $\chi$
\begin{equation}
J(\chi)=15+516\chi^2\left(\ln
\frac{\chi}{\sqrt{3}}-C\right)+1893\chi^2+\ldots \simeq
15\left[1-34.4\chi^2\left(\ln\frac{1}{\chi}-2.542\right)\right]
\label{a.16}
\end{equation}
In the case $\chi \gg 1$ it is convenient to present the integral
Eq.(\ref{a.14}) in the form
\begin{equation}
J(\chi)=\frac{i}{2}\int_{\lambda-i\infty}^{\lambda+i\infty}
\frac{\chi^{2s}}{3^s} \Gamma(1-s)\Gamma(1+3s)R(s)\frac{ds}{\sin\pi
s}, \quad -\frac{1}{3} < \lambda <0
\label{a.17}
\end{equation}
Closing the integration contour on the left one obtains the series
over the inverse powers of $\chi$
\begin{eqnarray}
&&J(\chi)=\frac{58\pi \Gamma(1/3)}{81\cdot 3^{1/6}\chi^{2/3}}+
\frac{628\pi3^{1/6} \Gamma(2/3)}{243 \chi^{4/3}}-\frac{13}{\chi^2}
\left(\ln \chi-\frac{1}{2}\ln 3-C+\frac{57}{52}\right)
\nonumber \\
&&+ \frac{188\pi \Gamma(1/3)}{81\cdot 3^{1/6}\chi^{8/3}}+\dots .
\label{a.18}
\end{eqnarray}

\newpage

{\bf Figure captions}

{\bf Fig.1} The inverse radiation length in tungsten, axis $<111>$
at different temperatures T vs the electron initial energy. Curves 1
and 4 are the total effect:
$L^{cr}(\varepsilon)^{-1}=I(\varepsilon)/\varepsilon$ Eq.(\ref{17})
for T=293 K and T=100 K correspondingly, the curves 2 and 5 give the
coherent contribution $I^F(\varepsilon)/\varepsilon$ Eq.(\ref{19}),
the curves 3 and 6 give the incoherent contribution
$I^{inc}(\varepsilon)/\varepsilon$ Eq.(\ref{21}) at corresponding
temperatures T.

{\bf Fig.2}

Enhancement (the ratio $L^{BM}/L^{ef}$) in tungsten, axis $<111>$,
T=293 K. The curve 1 is for the target with thickness $l=200~\mu m$,
where the energy loss was taken into account (according with
Eq.(\ref{18e})). The curve 2 is for a considerably more thinner
target, where one can neglect the energy loss ($L^{ef} \rightarrow
L^{cr}$). The data are from \cite{KMU}.

{\bf Fig.3}

The relative contribution of the LPM effect $\Delta$ (per cent) in
tungsten, axis $<111>$. Curve 1 is for T=100 K and curve 2 is for
T=293 K.

\newpage
\begin{table}
\begin{center}
{\sc Table 1}~ {Parameters of radiation process of the tungsten
crystal, axis $<111>$ for two temperatures T}
\end{center}
\begin{center}
\begin{tabular}{*{10}{|c}|}
\hline T(K)&$V_0$(eV)&$x_0$&$\eta_1$&$\eta$&
$\varepsilon_0$(GeV)&$\varepsilon_t$(GeV)&$\varepsilon_s$(GeV)&$\varepsilon_m$(GeV)&$h$ \\
\hline 293&413&39.7&0.108&0.115&7.43&0.76&34.8&14.4&0.348\\
\hline 100&355&35.7&0.0401&0.0313&3.06&0.35&43.1&8.10&0.612\\
\hline
\end{tabular}
\end{center}
\end{table}

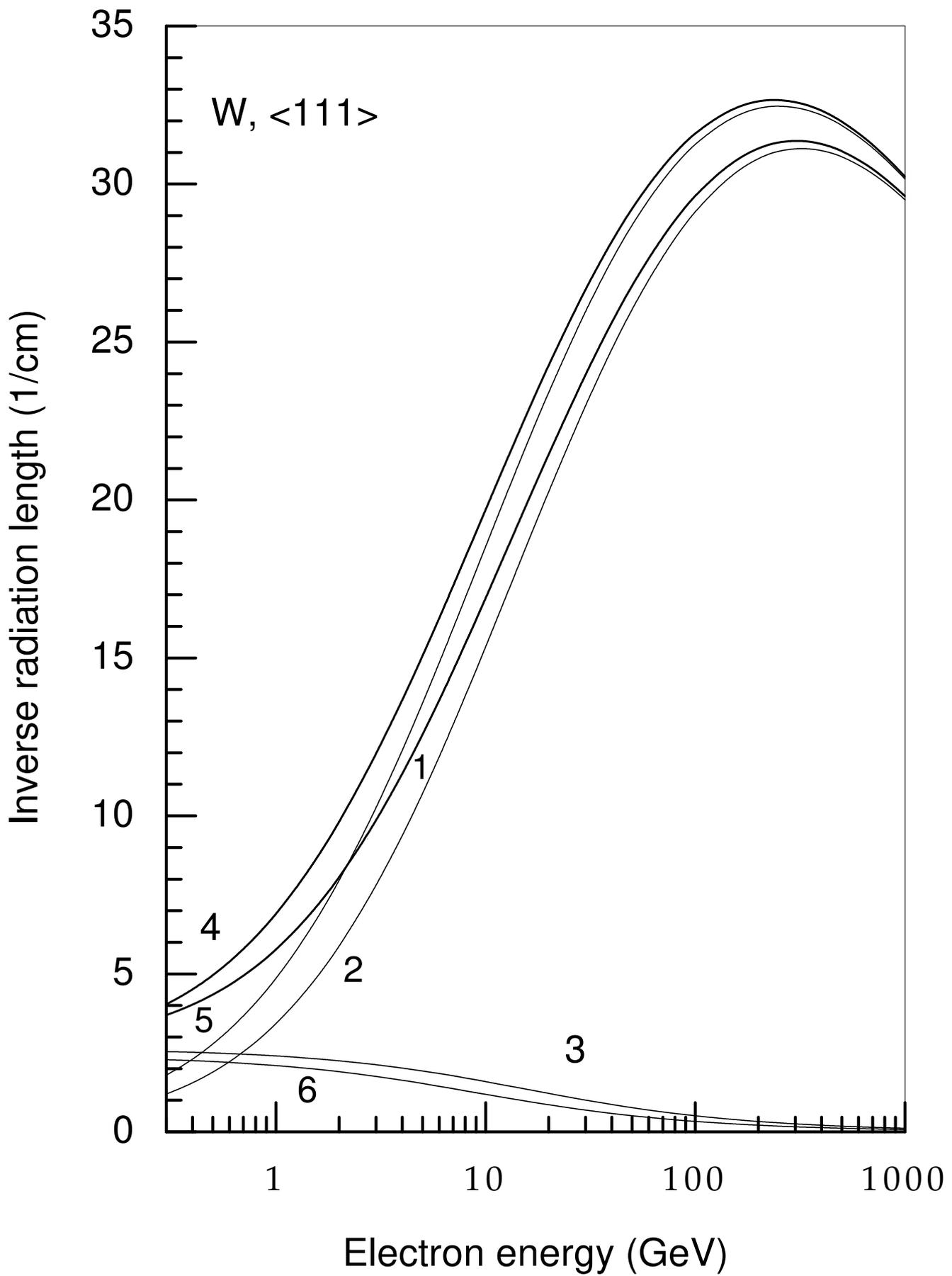
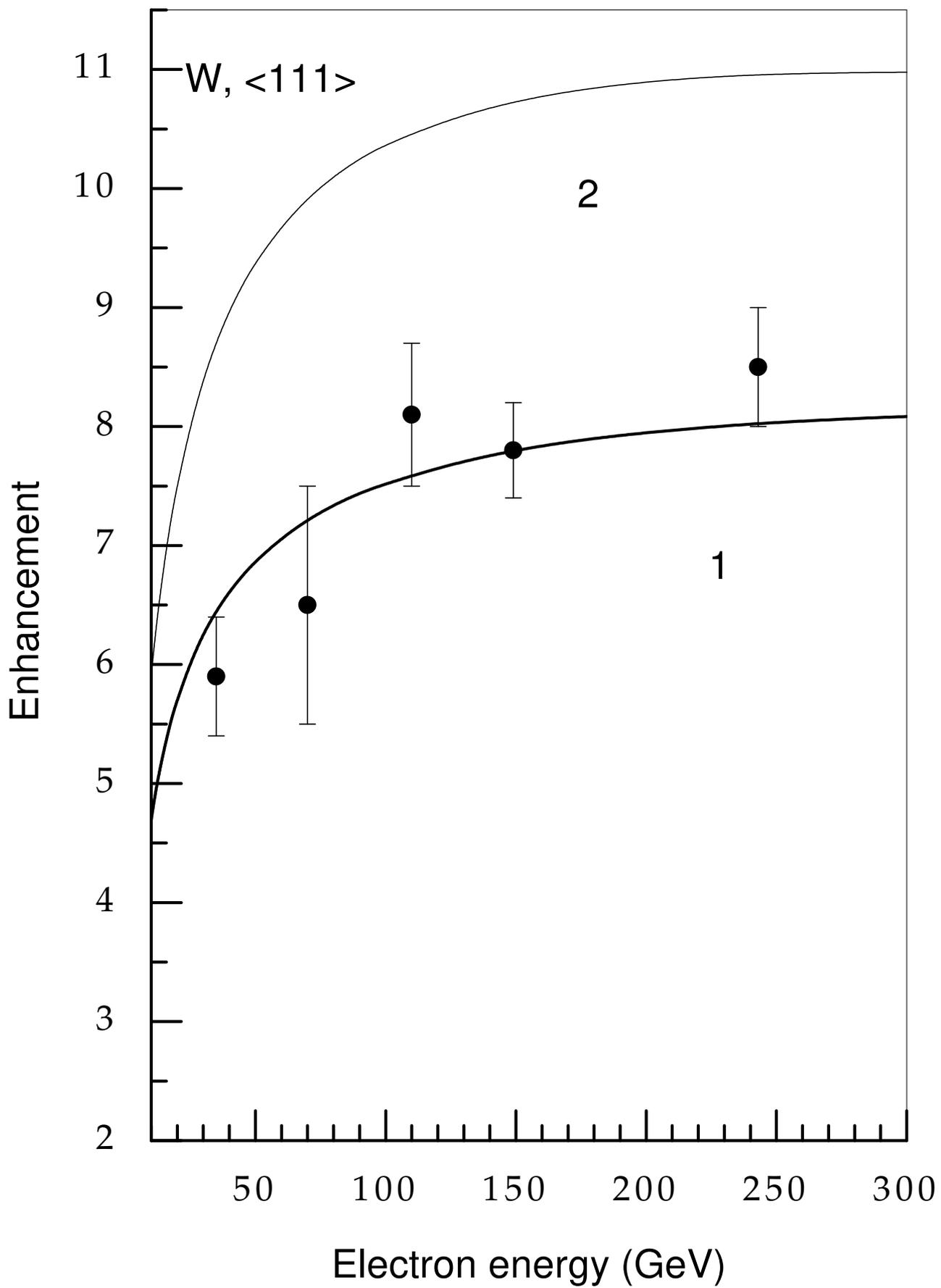
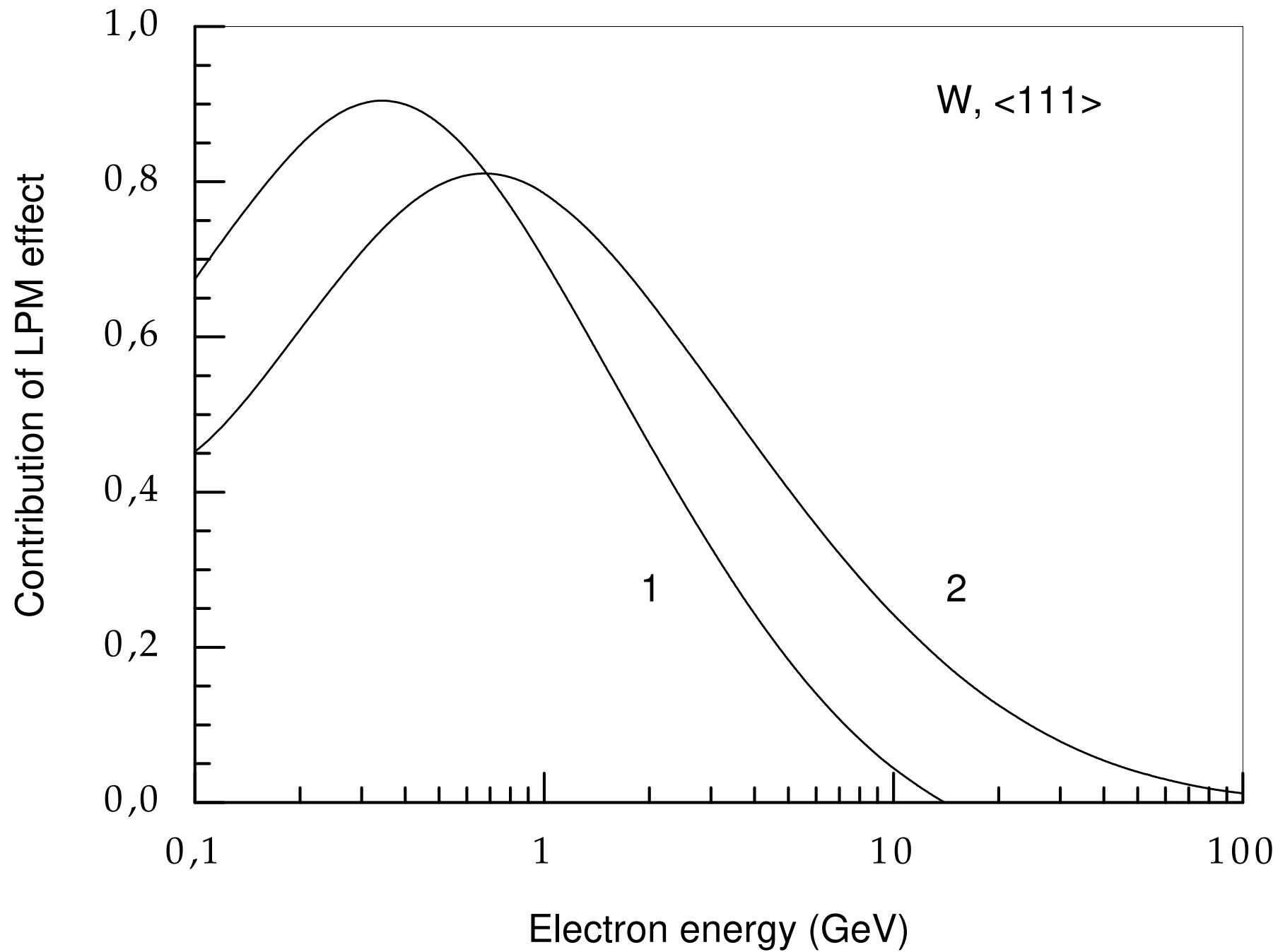
\end{document}